\documentclass[twocolumn,aps,prl,showpacs,superscriptaddress,twocolumn]{revtex4-1}
\usepackage{lmodern}

\usepackage[latin9]{inputenc}
\usepackage{amsmath}
\usepackage{amssymb}
\usepackage{graphicx}
\usepackage{epstopdf}
\usepackage{color}
\usepackage{enumerate}
\usepackage{ulem}
\usepackage{natbib}
\usepackage{caption}
\usepackage{soul,color}
\definecolor{orange}{RGB}{255,125,0}

\begin{document}
	
	\bibliographystyle{apsrev}

	\title{Limited role of vortices in transport in highly disordered superconductors near $B_{c2}$}

	\author{A. Doron}
	\email{adam.doron@weizmann.ac.il; Corresponding author}
	\affiliation{Department of Condensed Matter Physics, The Weizmann Institute of Science, Rehovot 7610001, Israel.}
	
	\author{T. Levinson}
	\affiliation{Department of Condensed Matter Physics, The Weizmann Institute of Science, Rehovot 7610001, Israel.}

	\author{F. Gorniaczyk}
	\affiliation{Department of Condensed Matter Physics, The Weizmann Institute of Science, Rehovot 7610001, Israel.}	

	\author{I. Tamir}
	\affiliation{Department of Condensed Matter Physics, The Weizmann Institute of Science, Rehovot 7610001, Israel.}
	\affiliation{Present Address: Fachbereich Physik, Freie Universit\"{a}t Berlin, 14195 Berlin, Germany.}
	
	\author{D. Shahar}
	\affiliation{Department of Condensed Matter Physics, The Weizmann Institute of Science, Rehovot 7610001, Israel.}

	\begin{abstract}
At finite temperatures and magnetic fields,  type-II superconductors in the mixed state have a non-zero resistance that is overwhelmingly associated with vortex motion.
In this work we study amorphous indium oxide films, which are thicker than the superconducting coherence length, and show that near $B_{c2}$ their resistance in the presence of perpendicular and in-plane magnetic fields becomes almost isotropic. Up to a linear rescaling of the magnetic fields both the equilibrium resistance as well as the non-equilibrium current-voltage characteristics are insensitive to  magnetic field orientation suggesting that, for our superconductors, there is no fundamental difference in transport between perpendicular and in-plane magnetic fields.
Additionally we show that this near-isotropic behavior extends to the insulating phase of amorphous indium oxide films of larger disorder strength that undergo a magnetic field driven superconductor-insulator transition. This near-isotropic behavior raises questions regarding the role of vortices in transport and the origin of resistance in thin-film superconductors.

	\end{abstract}
	
	\maketitle


\subsection{Introduction}	

Driving a transport current ($I$) through a thin film type-II superconductor in the mixed state results in finite dissipation and therefore finite resistance ($R$). 
This finite $R$ is overwhelmingly associated with vortex motion \cite{tinkham,bardeen1965theory,tinkham1964viscous,anderson1962theory,anderson1964hard,larkin1979pinning,feigel1990pinning,blatter1994vortices} where dissipation increases with vortex velocity which, in turn, is governed by an interplay between the Lorentz force, the vortex pinning force and a viscous damping of vortex motion \cite{bardeen1965theory,blatter1994vortices}.

In thin films the large discrepancy between thickness ($t$) and other sample dimensions (which we denote as $L$, typically $L/t\approx 10^{3}-10^{6}$) leads to an anisotropic response to perpendicular and parallel magnetic fields ($B_{\perp}$ and $B_{||}$) even in an intrinsically isotropic material. This anisotropic response can be understood by comparing the lengthscales of the problem that affect the vortex pinning force. If $t$ is smaller than the coherence length, $\xi$, which is the radius of a vortex (normal) core, vortices can penetrate the sample under the application of $B_{\perp}$ but not $B_{||}$ resulting in a pronounced anisotropy (some examples are high-$T_{c}$ superconductors \cite{palstra1989critical}, twisted bi-layer graphene \cite{cao2018unconventional} and other thin-films \cite{ruggiero1980superconductivity}). 

In films of intermediate thicknesses, where $L\gg t\gtrsim \xi$, vortices can penetrate both under $B_{\perp}$ and $B_{||}$ \cite{FootInPlaneVortices} but their contribution to $R$ is expected to be significantly different. This is because the vortex pinning force is proportional to the vortex length (assuming vortices have a finite elasticity \cite{larkin1979pinning,blatter1994vortices}) which are $t$ and $L$ in $B_{\perp}$ and $B_{||}$ respectively. In addition, the characteristic distance over which vortices logarithmically interact (setting the size of vortex bundles that are collectively pinned \cite{blatter1994vortices}) is the penetration depth which in $B_{\perp}$ is rescaled from its bulk value $\lambda$ to $\lambda_{\perp}=\lambda^{2}/t$ \cite{tinkham}. As $\lambda\gg t$ in our type-II superconductor,  $\lambda_{\perp}\gg \lambda$. 

Recently, we studied the critical current ($I_{c}$) in a:InO films at low $T$'s and high $B$'s near $B_{c2}$ \cite{DoronCriticalCurrents} and showed that $I_{c}$ is consistent with a thermal bi-stability.
To study the contribution of vortex motion to $I_{c}$ we measured $I-V$'s in $B_{||}$ for two different angles between the source-drain current and $B_{||}$ ($\varphi$): $\varphi=45^{\circ}$ and $\varphi=0^{\circ}$. Although the Lorentz force acting on vortices is $\propto \sin{\varphi}$, the measured $I-V$'s are independent of $\varphi$. This result suggests that $I_{c}$ is not a result of vortex de-pinning.

The anisotropy in a:InO films has been previously studied for samples of various disorder strengths by adjusting the angle between $B$ and the sample plane ($\theta$) \cite{gantmakher2000observation,paalanenprl69,johansson2011angular,shammass2012superconducting}.
Experimental studies \cite{johansson2011angular,shammass2012superconducting} showed that at low $B$'s there is a pronounced anisotropy on both sides of the disorder driven superconductor-insulator transition (SIT), where in samples that are superconducting (at $B=0$) $R$ initially scales with the orbital component of $B$ ($B_{\perp}=B\sin\theta$) but at higher $B$'s the anisotropy decreases and $R$ becomes isotropic. This was demonstrated for all samples regardless of whether they were superconducting or insulating at $B=0$.
This behavior was explained in Ref.~\cite{porat2015magnetoresistance} using a percolation model which assumes that the sample is composed of superconducting islands. The anisotropic behavior is due to orbital effects that reduce the coherent coupling between superconducting islands and the isotropic behavior at high $B$'s is a result of a Zeeman field that causes the collapse of superconducting islands.


The goal of this work we are reporting here is to characterize the anisotropy in the resistance response of  superconducting films of various thicknesses. We will show that: 1. Our superconductors are anisotropic at low $B$'s and turn isotropic near $B_{c2}$ (as reported previously \cite{johansson2011angular}) 2. We identified matching pairs of $B_{\perp}$ and $B_{||}$ where both equilibrium $R$ and the corresponding non-equilibrium $V-I$'s are identical, within experimental error. 3. Using a simple linear relation we can map all transport properties between $B_{||}$ and $B_{\perp}$ and 4. We demonstrate that a similar linear mapping between $B_{\perp}$ and $B_{||}$ applies to both equilibrium and non-equilibrium transport properties also in the $B$-driven insulating phase of a:InO films of higher disorder at $B$'s below the magneto-resistance peak.

\subsection{Experimental}	
The data of superconducting samples presented in this study were obtained from five a:InO films of various thicknesses t=22, 26, 57, 100 and 280 nm. Each sample was thermally annealed post deposition to a room-$T$  
$\rho$ of 4 $\pm 0.2 $ m$ \Omega \cdot$cm, which places them in the relatively low-disorder range of a:InO.
All measurements were performed using an Oxford instruments Kelvinox dilution refrigerator with a base $T$ of 10mK, equipped with a z-axis magnet. 
In order to control the angle between $B$ and the sample plane the samples were mounted on a probe with a rotating head.
While measuring in the superconducting phase, all measurement lines where filtered using room-$T$ RC filters with a cutoff frequency of 200 KHz. 

To facilitate a meaningful comparison among our films of various thickness, we consider our results in terms of intensive parameters: current density ($J=I/wt$), electric field ($E= V/l$), resistivity ($\rho=Rwt/l$) and differential resistivity ($dE/dJ$) where $l$ and $w$ are the sample's length and width and $V$ is the voltage.

\section{Results and analysis}	

\begin{figure}
	\centering
	\includegraphics [height=6cm] {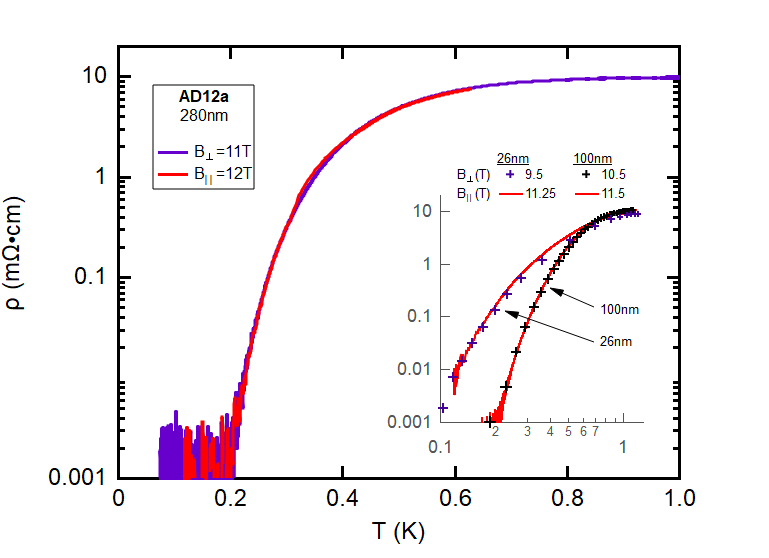}
	\caption{{\bf Equilibrium $\boldsymbol{\rho(T)}$ - $\boldsymbol{B_{\perp}}$ and $\boldsymbol{B_{||}}$.} 
		 Main frame: $\rho$ vs. $T$  for the 280nm thick film at $B_{\perp}=11$T (purple) and $B_{||}=12$T (red).
		 inset: $\rho$ vs. $T$ of the 100nm thick film at $B_{\perp}=10.5$T (black crosses) and $B_{||}=11.5$T (red dashed line) and for the 26nm thick film at $B_{\perp}=9.5$T (purple crosses) and $B_{||}=11.25$T (red continuous line).
	}			
	\label{RoleOfVortices:Figure1}
\end{figure}

We begin by showing that we can identify pairs of $B$'s, $B_{\perp}$-$B_{||}$, for which the transport characteristics are virtually indistinguishable. In Fig.~\ref{RoleOfVortices:Figure1} we plot $\rho$ vs. $T$ of the 280nm thick film at $B_{||}=12$T (red) and $B_{\perp}=11$T (purple). In this pair of $B$'s we find that the $\rho(T)$ curves are identical, within error, for both $B$ orientations. We refer to such pairs of $B_{\perp}$ and $B_{||}$, having the same $\rho(T)$, as matching-$B$'s. We could find such matching-$B$'s virtually for our entire measurement range. In the inset of Fig.~\ref{RoleOfVortices:Figure1} we plot similarly matching-$B$'s obtained from the 100nm and 26nm films.


\begin{figure}
		\centering
		\includegraphics [height=6 cm] {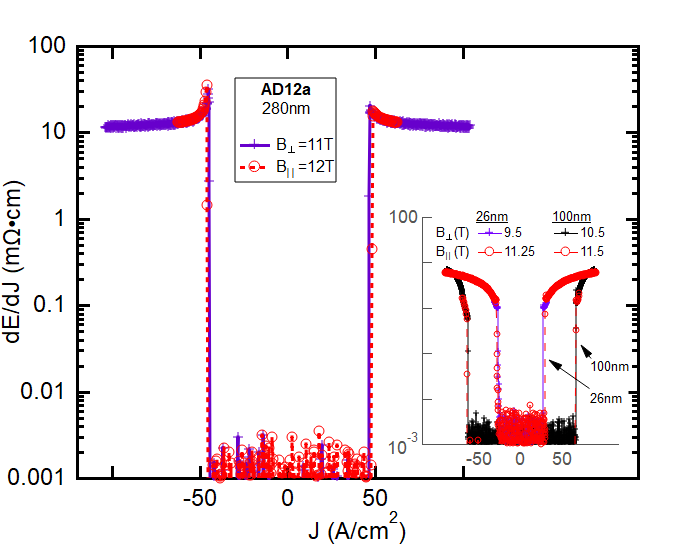}
	\caption{{\bf Non-equilibrium $\boldsymbol{\frac{dE}{dJ}}$ vs. $\boldsymbol{J}$ - $\boldsymbol{B_{\perp}}$ and $\boldsymbol{B_{||}}$.} 
	Main frame: $\frac{dE}{dJ}$ vs. $J$ of the 280nm thick film at $B_{\perp}=11$T (purple crosses) and $B_{||}=12$T (red circles).
	inset:$\frac{dE}{dJ}$ vs. $J$ of the 100nm thick film at $B_{\perp}=10.5$T (black crosses) and $B_{||}=11.5$T (red circles) and for the 26nm thick film at $B_{\perp}=9.5$T (purple crosses) and $B_{||}=11.25$T (red circles).
}				
	\label{RoleOfVortices:Figure2}
\end{figure}	

Next we show that the correspondence within matching $B$'s extends beyond Ohmic transport and equally applies to the critical current of superconductivity, and beyond. In Fig.~\ref{RoleOfVortices:Figure2} we plot $\frac{dE}{dJ}$ vs. $J$ of the 280nm thick film at $B_{\perp}=11$T (purple) and $B_{||}=12$T (red), the same matching-$B$'s as in Fig.~\ref{RoleOfVortices:Figure1}.
In the inset of Fig.~\ref{RoleOfVortices:Figure2} we plot $\frac{dE}{dJ}$ vs. $J$ for 26nm and 100nm thick films at the same matching $B$'s as those in the inset of Fig.~\ref{RoleOfVortices:Figure1}.

The data presented in Figs. \ref{RoleOfVortices:Figure1} and \ref{RoleOfVortices:Figure2} portray the main point of this work: there is no fundamental difference in the transport properties of our superconductors between $B_{||}$ and $B_{\perp}$ and our samples, despite being thin-films, are in effect isotropic (as evident by the closeness of the $|B|$ values in matching-$B$'s. The minor $B$ differences in matching-$B$'s are discussed below). This holds for equilibrium (Fig.~\ref{RoleOfVortices:Figure1}) as well as non-equilibrium (Fig.~\ref{RoleOfVortices:Figure2}) transport.

This near-isotropic transport at high $B$'s deserves further consideration. In type-II superconductors the finite $\rho$ measured below the superconducting critical $T$ is typically associated with vortex motion. In our samples $\xi$ ($\approx 5$nm \cite{sacepe2015high}) is smaller than the thickness therefore vortices can penetrate through the sample plane. As the vortex pinning-energy scales with its length \cite{blatter1994vortices}, which changes by 3-4 orders of magnitude between the $B$-orientations, one would expect $\rho$ to display a significant anisotropy. The measured (near-)isotropic behavior suggests that the origin of the finite $\rho$ measured at low $T$'s is not likely due to vortex motion.

\begin{figure} [t!]
	\centering
	\includegraphics [height=4.8 cm] {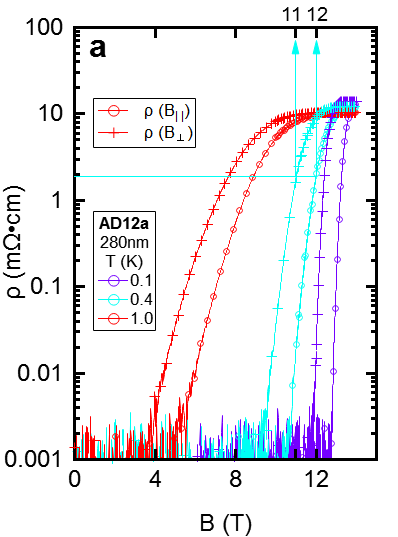}
	\includegraphics [height=4.8 cm] {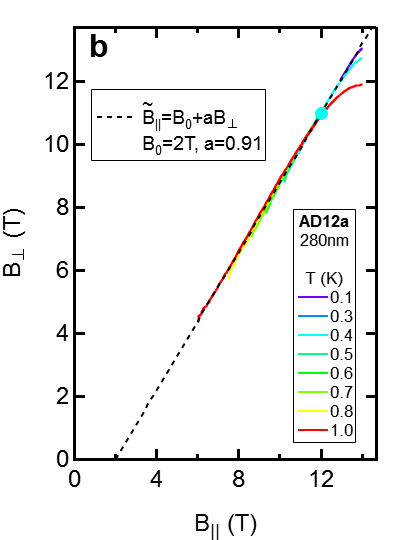}
	\includegraphics [height=4.8 cm] {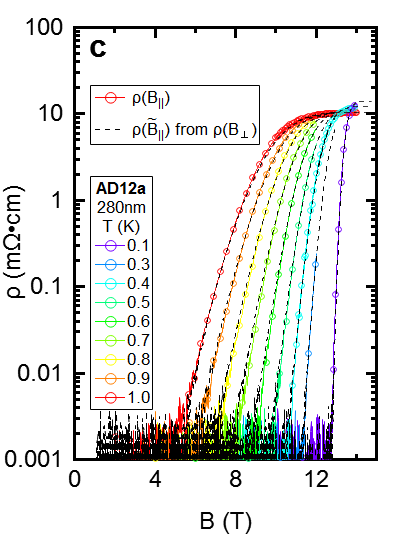}		
	\includegraphics [height=4.8 cm] {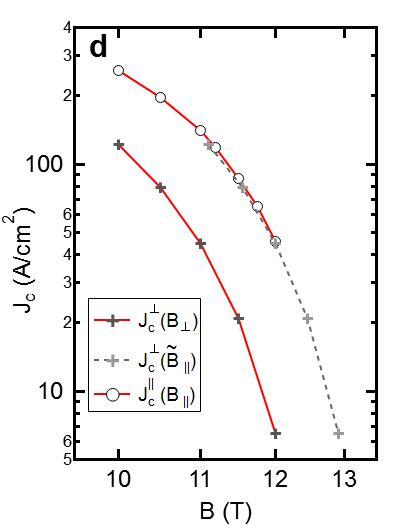}	
	\caption{{\bf Linear mapping between $\boldsymbol{B_{\perp}}$ and $\boldsymbol{B_{||}}$.}
		(a) $\rho(B_{\perp})$ (crosses) and $\rho(B_{||})$ (circles) at $T=0.1$K (purple), $0.4$K (light blue) and $1$K (red).
		(b) Continuous mapping between $\rho$ at $B_{||}$ and $B_{\perp}$. Each point on the $B_{||}-B_{\perp}$ plane satisfies $\rho(B_{||},T)=\rho(B_{\perp},T)$. The color-coding marks different $T$'s and the dashed black line is a linear fit.
		(c) Measured $\rho(B_{||})$ (colors) overlapped with rescaled $\rho(\tilde{B}_{||})$ (black). 
		(d) $J^{\perp}_{c}$ vs. $B_{\perp}$ (black crosses connected by a red line) and $J^{||}_{c}$ vs. $B_{||}$  (circles connected by a red line) of the 280nm thick film. Gray crosses connected by a dashed gray line are $J^{\perp}_{c}$ displayed vs. a rescaled abscissa $\tilde{B}_{||}$ according to the linear mapping of (b) and Eq.~\ref{RoleOfVortices:eLinMap}.
	}			
	\label{RoleOfVortices:Figure3}
\end{figure}

To gain further insight into this $B$-orientation dependence, we inspect the shift in $B$ among the matching $B$'s.
In Fig.~\ref{RoleOfVortices:Figure3}a we plot $\rho$ vs. $B_{||}$ and $B_{\perp}$ of the 280nm thick sample at various $T$'s. 
To identify the law for matching $B$'s we perform a continuous mapping between $B_{||}$ and $B_{\perp}$ by finding $B$-pairs that satisfy $\rho(B_{||},T)=\rho(B_{\perp},T)$. For example, the horizontal light-blue line in Fig.~\ref{RoleOfVortices:Figure3}a corresponds to $\rho =1.9$~m$\Omega \cdot$cm and the vertical light blue arrows mark its intersect with $\rho(B,T=0.4K)$ in both $B$-orientations. In this example $\rho(B_{||}=12$T$,T=0.4$K$)=\rho(B_{\perp}=11$T$,T=0.4$K$)$. 
In Fig.~\ref{RoleOfVortices:Figure3}b we plot the result of this continuous mapping on a $B_{||}-B_{\perp}$ plane, the $B_{||}=12$T, $B_{\perp}=11$T matching-$B$'s from the example above is marked by a light blue circle. The continuous lines are obtained by continuously varying $\rho$ and satisfying the condition $\rho(B_{||},T)=\rho(B_{\perp},T)$ and the colors correspond to performing the mapping at different $T$'s.
The dashed black line marks the following linear mapping between equilibrium transport at the different $B$ orientations
\begin{equation}
	\tilde{B}_{||}=B_{0}+aB_{\perp}
\label{RoleOfVortices:eLinMap}
\end{equation}
where for the 280nm thick sample $B_{0}=2$T and $a=0.91$ and $\tilde{B}_{||}$ ($\tilde{B}_{\perp}$) stands for a measurement at $B_{\perp}$ ($B_{ ||}$) rescaled according to Eq.~\ref{RoleOfVortices:eLinMap}.

The quality of this linear mapping between both $B$-orientations is demonstrated in Fig.~\ref{RoleOfVortices:Figure3}c where circles connected by colored lines mark $\rho(B_{||})$ (colors mark different isotherms) and the dashed black lines mark $\rho(\tilde{B}_{||})$ at the corresponding $T$'s, i.e. the result of applying the mapping of Eq.~\ref{RoleOfVortices:eLinMap} on $\rho (B_{\perp})$ measured data. 

\begin{figure*}
	\centering
	\includegraphics [height=5.5 cm] {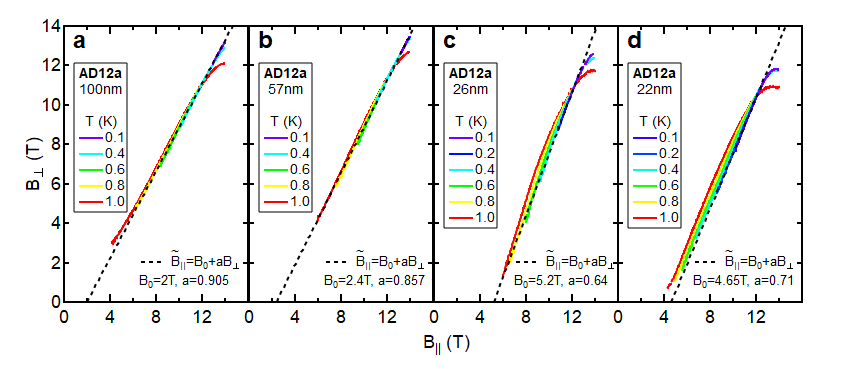}
	\caption{{\bf $\boldsymbol{B_{\perp}}$ and $\boldsymbol{B_{||}}$ mapping for a:InO films of various thicknesses} 
		(a)-(d)	Continuous mapping between $\rho$ at $B_{||}$ and $B_{\perp}$ as in Fig.~\ref{RoleOfVortices:Figure2}a for films of thicknesses 100nm (a), 57nm (b), 26nm (c) and 22nm (d) where each point on the $B_{||}\;-\;B_{\perp}$ plane marks $\rho(B_{||},T)=\rho(B_{\perp},T)$, the color-coding marks different $T$'s and the dashed black line is a linear fit to the data.
	}			
	\label{RoleOfVortices:Figure4}
\end{figure*}

To test whether the mapping described in Eq.~\ref{RoleOfVortices:eLinMap} and extracted from Ohmic measurements also applies to non-equilibrium transport we study whether this mapping can account for the anisotropy in critical current density ($J_{c}$) \cite{FootnoteTrapping}.  In Fig.~\ref{RoleOfVortices:Figure3}d we plot $J_{c}$ measured at $B_{\perp}$ ($J_{c}^{\perp}$, black crosses connected by a red line) and $J_{c}$ measured at $B_{||}$ ($J_{c}^{||}$, circles connected by a red line) vs. the magnitude of $B$. Gray crosses connected via a dashed gray line mark $J^{\perp}_{c}(\tilde{B}_{||})$, namely $J_{c}$ measured in $B_{\perp}$ where the abscissa was rescaled according to Eq.~\ref{RoleOfVortices:eLinMap}.
The collapse of $J^{\perp}_{c}(\tilde{B}_{||})$ and $J_{c}^{||}(B_{||})$ suggests that the linear mapping also holds for non-equilibrium transport. 
This result is consistent with the conclusions of Ref.~\cite{DoronCriticalCurrents} where we showed that the critical currents are determined entirely by the Ohmic $R(T)$.

We repeated the process of mapping between the equilibrium $\rho(B_{\perp})$ and $\rho(B_{||})$ described above for four more samples of different thicknesses: 100nm, 57nm, 26nm and 22nm (See supplemental material \cite{SupplementalRoleOfVortices} Sec.~S1 for $\rho(B_{\perp})$ and $\rho(B_{||})$ of all samples). The results of these mappings are plotted in Figs.~\ref{RoleOfVortices:Figure4}a-d (see Table.~1 in the supplemental material \cite{SupplementalRoleOfVortices} for a summary of the mapping parameters).
The mapping parameters $B_{0}$ and $a$ of the 100nm and 57nm thick films (Fig.~\ref{RoleOfVortices:Figure4}a-b) are similar to those of the thicker 280nm thick film (Fig.~\ref{RoleOfVortices:Figure3}b).
For 26nm and 22nm thick films the mappings performed at different $T$'s did not fully converge as a whole into a single curve. We chose to perform the linear fit such that it best describes the low-$T$ range of the mapping. 
The resulting $B_{0}$ and $a$ do differ from that of the thicker samples.
In the discussion section we consider reasons for the differences in mappings between thinner and thicker films.
In Sec.~S2 of the supplemental material \cite{SupplementalRoleOfVortices} we show that, similarly to Fig.~\ref{RoleOfVortices:Figure3}d, the equilibrium mappings displayed in Fig.~\ref{RoleOfVortices:Figure4} also maps the critical currents of both $B$-orientations.
\subsection{Discussion}	

\subsubsection*{Role of vortices in our type-II superconductor}

The main conclusion from the results presented above is that there is no fundamental difference between transport at $B_{\perp}$ and $B_{||}$, especially near $B_{c2}$ where samples become isotropic \cite{johansson2011angular,shammass2012superconducting,porat2015magnetoresistance} (see supplemental material \cite{SupplementalRoleOfVortices} Fig.~S2 for a study of vanishing anisotropy near $B_{c2}$ in our superconducting samples). 
As vortices and pinning forces are expected to be drastically different between these $B$-orientations, our results question the role of vortices in transport of highly disordered superconductors near $B_{c2}$.
Adding to that the aforementioned result reported in Ref.~\cite{DoronCriticalCurrents}, that  $J_{c}^{||}$ is independent of the angle $\varphi$ between $J$ and $B_{||}$, suggests that the finite $R$ and $J_{c}$ measured in our type-II superconductor at a finite $T$ is caused by a non-vortex mechanism and raises the interesting question of what actually causes $R$ in highly disordered superconductors near $B_{c2}$.


\subsubsection*{Mapping between $\boldsymbol{B_{\perp}}$ and $\boldsymbol{B_{||}}$ near the SIT}
Up to now we discussed the similarity in transport between $B_{\perp}$ and $B_{||}$ only in relatively low disordered a:InO films that do not have a prominent insulating phase.
a:InO films of higher disorder strength undergo a SIT \cite{goldmanpt51,physupekhi,sondhirmp} that can be driven by various parameters \cite{HebardPrl,kapitulnikprl74,BaturinaJETP,Shaharprb,goldmanprl94,haviprl62,goldmanpt51}. It turns out that the above analysis is also relevant to higher disordered samples near the $B$ driven SIT \cite{HebardPrl,kapitulnikprl74,BaturinaJETP}. 

\begin{figure*}
	\centering
	\includegraphics [height=5.5 cm] {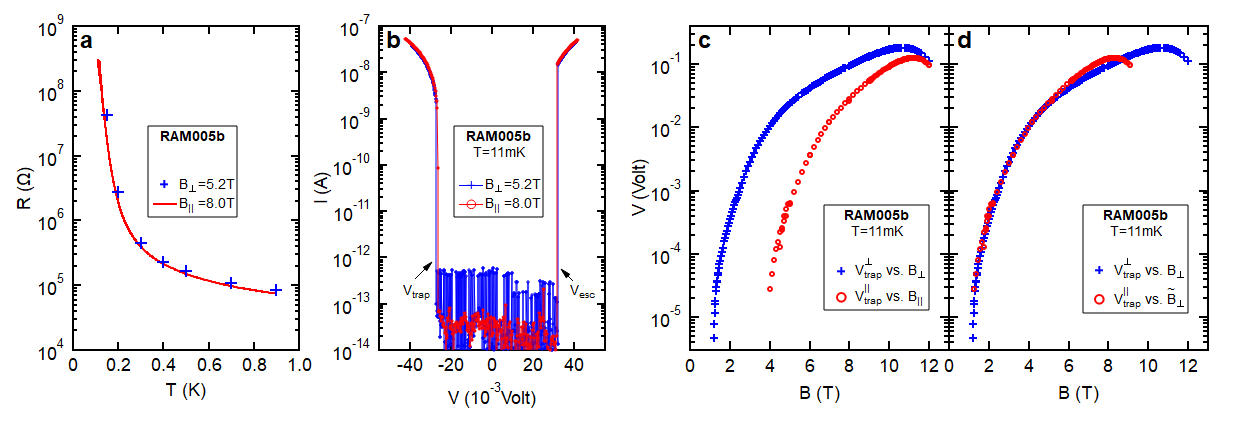}
	\caption{{\bf Mapping in the insulating phase of a:InO.} 
		$R(T)$ (a) and 11mK $I-V$'s (b) of sample RAM005b measured at $B_{\perp}=$5.2T (blue) and $B_{||}=$8T (red) in the $B$ driven insulating phase.
		(c) Blue crosses and red circles mark $V_{trap}^{\perp}(B_{\perp})$ and $V_{trap}^{||}(B_{||})$ at $T=11mK$.
		(d) $V_{trap}^{||}$ plotted vs. $\tilde{B}_{\perp}=(B_{||}-B_{0})/a$ instead of $B_{||}$ where $B_{0}=2.72$T and $a=1.02$ (the mapping between both $B$ orientations was extracted from equilibrium $R(B)$'s, see supplemental material \cite{SupplementalRoleOfVortices}).
	}			
	\label{RoleOfVortices:Figure5}
\end{figure*}

Below we consider sample RAM005b, a 30nm thick film that undergoes both $B_{\perp}$ and $B_{||}$ driven SIT's (see Sec.~S3 of the supplemental material \cite{SupplementalRoleOfVortices} for $R(B_{\perp})$ and $R(B_{||})$). 
In Fig.~\ref{RoleOfVortices:Figure5}a we plot $R(T)$ at $B_{\perp}=$5.2T (blue crosses) and at $B_{||}=$8T (red line). In Fig.~\ref{RoleOfVortices:Figure5}b we plot $|I|$ vs. $V$ at the same matching-$B$'s at $T=$11mK. 
We repeated the same mapping protocol described above and found a linear relation between equilibrium $R(B_{\perp})$ and $R(B_{||})$ as in Eq.~\ref{RoleOfVortices:eLinMap} with $a\approx 1$ and $B_{0}\approx 2.7$T (see supplemental material \cite{SupplementalRoleOfVortices} for full details of the mapping). 

Next we apply the mapping to non-equilibrium transport. The non-equilibrium properties we map in the insulating phase are the threshold $V$'s, apparent in Fig.~\ref{RoleOfVortices:Figure5}b and marked as $V_{trap}$ and $V_{esc}$, where the $I-V$'s exhibit a large discontinuity in $I$ \cite{murthyprl} due to electron overheating \cite{borisprl,maozprl}.
In Fig.~\ref{RoleOfVortices:Figure5}c we plot $V_{trap}^{\perp}$ ($V_{trap}$ measured in $B_{\perp}$) vs. $B_{\perp}$, and $V_{trap}^{||}$ ($V_{trap}$ measured in $B_{||}$) vs. $B_{||}$. It is apparent that at low $B$'s $V_{trap}$ is anisotropic.
In Fig.~\ref{RoleOfVortices:Figure5}d we plot the same $V_{trap}$ but now we plot $V_{trap}^{||}$ vs. $\tilde{B}_{\perp}$ instead of vs. $B_{||}$.
Similar to the superconducting phase, the insulating phase also exhibits the apparent collapse of the threshold $V$'s (in the range $B_{||}<8$T and $B_{\perp}<5$T). 

The main difference between the mappings in both phases is in its regime of applicability. While in the superconducting phase the mapping applies for the whole $B$ range, including near $B_{c2}$ where $R$ becomes isotropic, in the insulating phase the mapping breaks at high $B$'s ($B_{||}>8$T and $B_{\perp}>5$T). We note that in the $B$ ranges where the mapping applies, the insulating phase of a:InO has an underlying superconducting nature where transport is carried mainly by Cooper-pairs that become spatially localized \cite{FeigAnnals,YonatanNat,GantmakherJETP,vallesprl103,BenjaminNat,sacepe2015high,craneprb751,paalanenprl69,breznay2016self,steiner2005superconductivity}. At higher $B$'s, beyond a well studied peak in the magnetoresistance \cite{paalanenprl69}, it is believed that superconductivity terminates locally and transport is carried mainly by quasi-particles \cite{paalanenprl69,breznay2016self,steiner2005superconductivity}. In that range in the insulator the mapping is no longer applicable.

\subsubsection*{Linear mapping between transport at $\boldsymbol{B_{\perp}}$ and $\boldsymbol{B_{||}}$}
The basis for the linear relation between $B_{\perp}$ and $B_{||}$, described by Eq.~\ref{RoleOfVortices:eLinMap}, is not clear yet. We summarize below all of our major observations that a future theoretical description should account for:
1. There is a linear mapping relating transport properties in $B_{\perp}$ and $B_{||}$ that equally applies to equilibrium and non-equilibrium situations.
2. The fit parameters $B_{0}$ and $a$ do not vary much between samples of thicknesses greater than 57nm, but these parameters are significantly different for the 22nm and 26nm thick films. This suggests that there is an important lengthscale which is greater than 26nm and smaller than 57nm. This lengthscale is unknown to us as for a:InO samples $\xi\sim 5$nm \cite{sacepe2015high}, $\lambda \sim 7\mu$m (see Sec.~S2 of the supplemental material of Ref.~\cite{DoronCriticalCurrents} which uses experimental results of Refs.~\cite{misra2013measurements,craneprb751}), the mean-free path is smaller than 1nm, the magnetic length corresponding to $B_{0}$ in samples thicker than 57nm is $\lesssim$19nm and in the 22nm and 26nm thick samples it is $\lesssim$12nm.
3. In the thinner 22nm and 26nm thick samples the mapping has some $T$-dependence.
4. The mapping also applies to the $B$-driven insulating phase of a:InO for $B$'s below the magneto-resistance peak.

One possible explanation for the isotropic behavior is that our samples are practically three dimensional. Following this logic one can account for the finite $B_{0}$ values as some geometric rescaling due to the large discrepancy between $t$ and the other sample dimensions that accounts for the finite nature of $t$. If that was indeed the case we would expect that as we approach the 3d limit, namely as $t$ increases, the mapping parameters would tend towards $a\to 1$ and $B_{0}\to 0$. Comparing these parameters of the 57nm, 100nm and 280nm thick films shows that $B_{0}\not\to 0$.

Another possible interpretation is that $B_{0}$ is related to a finite $H_{c1}$ of the sample where, at low $B_{||}$'s, there is a full Meissner effect. It was pointed to us \cite{MeirPrivate} that following this interpretation and using a slightly extended percolation approach than that of Ref.~\cite{porat2015magnetoresistance} can lead to the linear relation of Eq.~\ref{RoleOfVortices:eLinMap}. It is unclear to us whether this approach can also explain the linear mapping in the insulating phase.


In summary, the isotropic behavior detailed in this work suggests that the role of vortices in the mixed state of our highly disordered type-II superconductor is limited. We show that the rather weak measured anisotropy can be adjusted using a linear mapping between $B_{\perp}$ and $B_{||}$.

\begin{acknowledgments}
We are grateful to Y. Meir and K. Michaeli for fruitful discussions. 
	\paragraph{Funding}
This research was supported by The Israel Science Foundation (ISF Grant no. 556/17), the United States - Israel Binational Science Foundation (BSF Grant no. 2012210) and the Leona M. and Harry B. Helmsley Charitable Trust.
\end{acknowledgments}


\begin{thebibliography}{43}
\bibitem[{Foo({\natexlab{a}})}]{FootInPlaneVortices}
	\bibinfo{note}{In-plane vortices have their symmetry axis parallel to the
		sample plane.}
\bibitem[{Foo({\natexlab{b}})}]{FootnoteTrapping}
	\bibinfo{note}{$I_{c}$ has a small hysteresis. We swept $I$ from negative to
		positive values and define $I_{c}$ as that in the negative "trapping" side of
		the $V-I$'s}.
\end{thebibliography}

\end{document}


\title{Supplemental Material for Limited role of vortices in transport in highly disordered superconductors near $B_{c2}$}

	\author{A. Doron}
\email{adam.doron@weizmann.ac.il; Corresponding author}
\affiliation{Department of Condensed Matter Physics, The Weizmann Institute of Science, Rehovot 7610001, Israel.}

\author{T. Levinson}
\affiliation{Department of Condensed Matter Physics, The Weizmann Institute of Science, Rehovot 7610001, Israel.}

\author{F. Gorniaczyk}
\affiliation{Department of Condensed Matter Physics, The Weizmann Institute of Science, Rehovot 7610001, Israel.}	

\author{I. Tamir}
\affiliation{Department of Condensed Matter Physics, The Weizmann Institute of Science, Rehovot 7610001, Israel.}
\affiliation{Present Address: Fachbereich Physik, Freie Universit\"{a}t Berlin, 14195 Berlin, Germany.}

\author{D. Shahar}
\affiliation{Department of Condensed Matter Physics, The Weizmann Institute of Science, Rehovot 7610001, Israel.}

\maketitle


\section{$\boldsymbol{\rho}$ vs. $\boldsymbol{B_{\perp}}$ and vs. $\boldsymbol{B_{||}}$}
\begin{figure*}
	\centering
	\includegraphics [height=10 cm] {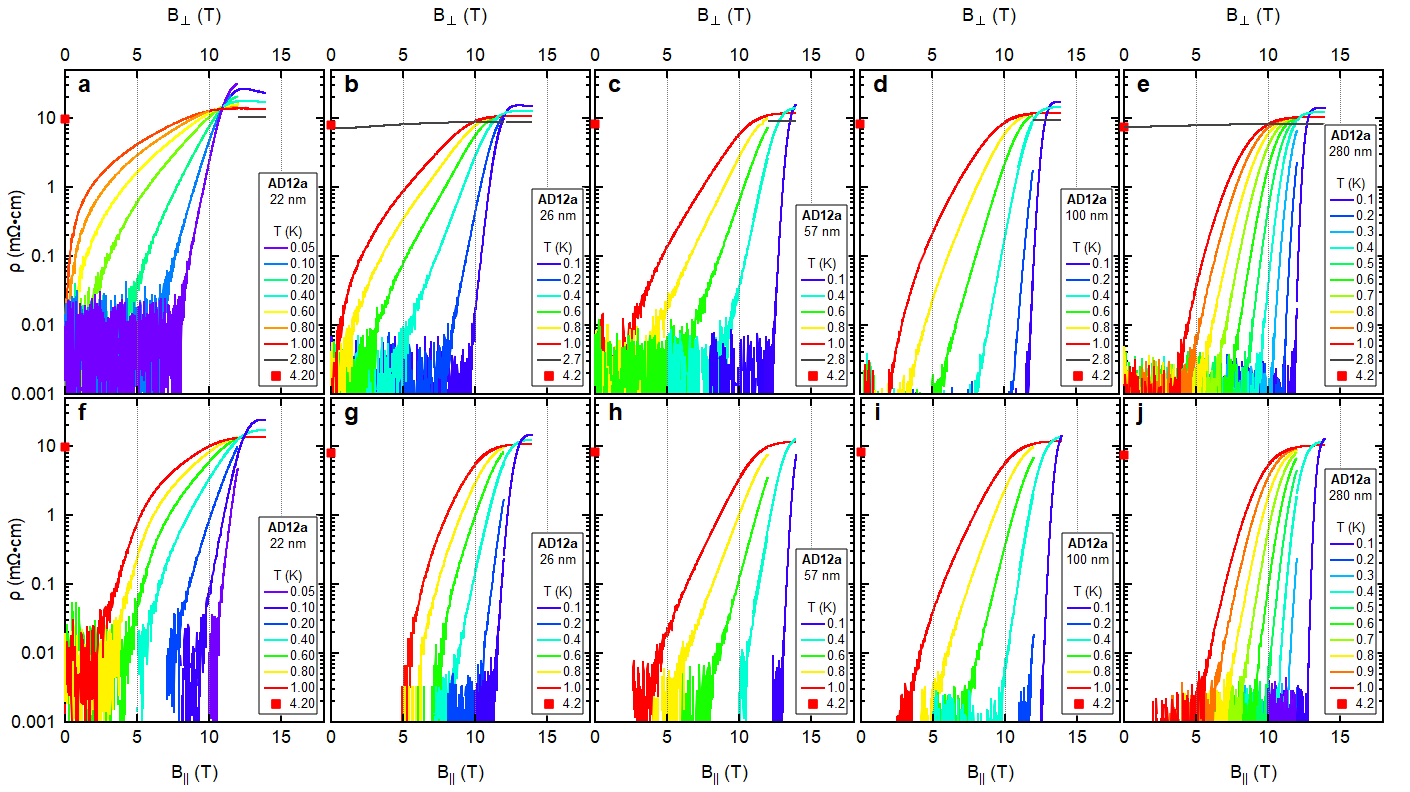}
	\caption{{\bf $\boldsymbol{\rho}$ vs. $\boldsymbol{B_{\perp}}$ and vs. $\boldsymbol{B_{||}}$.} 
		$\rho$ vs. $B_{\perp}$ of the 22nm (a) 26nm (b) 57nm (c) 100nm (d) and 280nm (e) thick films.\\
		$\rho$ vs. $B_{||}$ of the 22nm (f) 26nm (g) 57nm (h) 100nm (i) and 280nm (j) thick films.
	}			
	\label{RoleOfVortices:Supp:FigS1}
\end{figure*}

In the main-text we study five a:InO films of various thicknesses at $B_{\perp}$ and at $B_{||}$. 
In Figs.~\ref{RoleOfVortices:Supp:FigS1}a-e we plot $\rho$ vs. $B_{\perp}$ of these films at various $T$'s.
In Figs.~\ref{RoleOfVortices:Supp:FigS1}f-j we plot $\rho$ vs. $B_{||}$ of these films at various $T$'s.

\begin{figure*} [h!]
	\centering
	\includegraphics [height=10 cm] {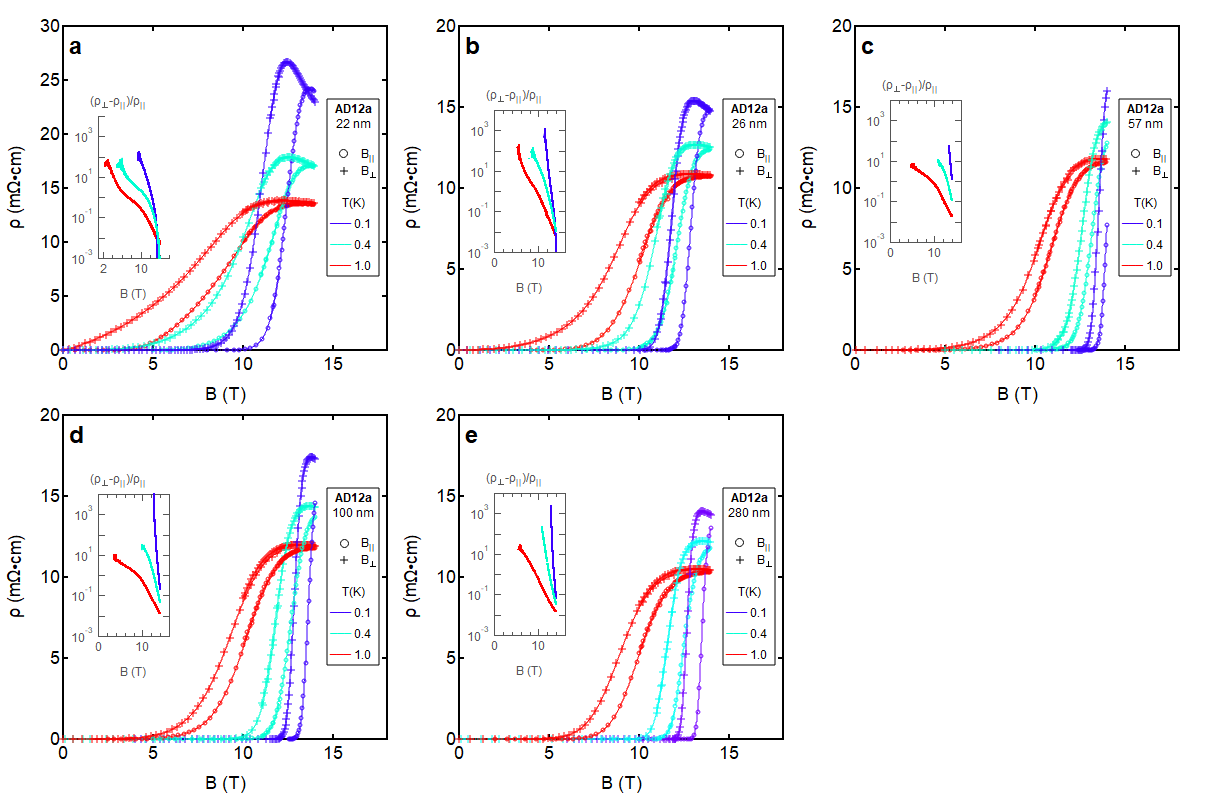}
	\caption{{\bf $\boldsymbol{\rho}$ vs. $\boldsymbol{B_{\perp}}$ and vs. $\boldsymbol{B_{||}}$.} 
		$\rho$ vs. $B_{\perp}$ (crosses) and vs. $B_{||}$ (circles) for 0.1K, 0.4K and 1K of the 22nm (a) 26nm (b) 57nm (c) 100nm (d) and 280nm (e) thick films. Insets: relative anisotropy $\frac{\rho(B_{\perp})-\rho(B_{||})}{\rho(B_{||})}$
	}			
	\label{RoleOfVortices:Supp:FigS2}
\end{figure*}

In the main-text we discuss the observation that the films become more isotropic with increasing $|B|$. In Figs.~\ref{RoleOfVortices:Supp:FigS2}a-e we plot $\rho$ vs. $B_{\perp}$ (crosses) and vs. $B_{||}$ (circles) measured at three different $T$'s (0.1K, 0.4K and 1K) of all films.
It can be seen that as $|B|$ increases the difference between both $B$ orientation decreases making the samples more isotropic.
In the two thinner films (22nm and 26nm), the lowest isotherm of 0.1K displays a reversed anisotropy at high $B$'s where $\rho(B_{||})>\rho(B_{\perp})$. A similar effect was previously reported in a:InO films [16] and its origin is not yet understood (we note that Ref.~[18] does account for this effect).

In the inset of these figures we plot the relative anisotropy defined as $\frac{\rho(B_{\perp})-\rho(B_{||})}{\rho(B_{||})}$ for $T=$0.1K, 0.4K and 1K.
It can be seen that the relative anisotropy decays exponentially while approaching $B_{c2}$ (and, as discussed above, can turn negative).

\section{Mapping of $\boldsymbol{I_{c}}$ between $\boldsymbol{B_{\perp}}$ and $\boldsymbol{B_{||}}$ in the superconducting phase}
\begin{figure*} 
	\centering
	\includegraphics [height=6 cm] {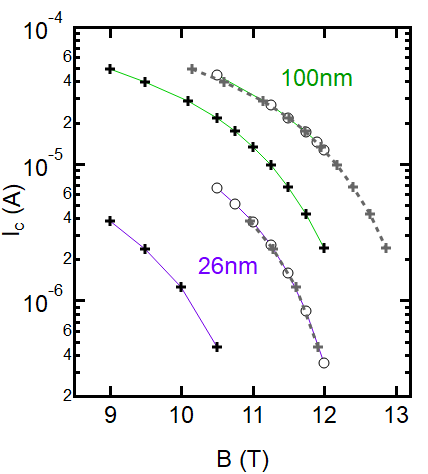}
	\caption{{\bf Mapping of non-equilibrium $\boldsymbol{I_{c}}$'s.} 
		Rescaling of $I_{c}$ for the 26nm (purple) and 100nm (green) thick films. $I^{\perp}_{c}$ vs. $B_{\perp}$ (black crosses connected by continuous lines) and $I^{||}_{c}$ vs. $B_{||}$  (circles connected by continuous lines). Gray crosses connected by dashed gray lines are $I^{\perp}_{c}$ displayed vs. a rescaled abscissa $\tilde{B}_{||}$ according to the linear mappings.
	}			
	\label{RoleOfVortices:Supp:FigS3}
\end{figure*}

In Fig.~3 of the main-text we show that for the 280nm thick sample the linear mapping described by Eq.~1 where the parameters $B_{0}$ and $a$ are extracted from the equilibrium zero bias $\rho$ also maps $J_{c}^{\perp}$ onto $J_{c}^{||}$.
In Fig.~4 of the main-text we show that similar linear mappings can also be extracted from zero bias $\rho$ of the 100nm, 57nm, 26nm and 22nm thick samples.
In Fig.~\ref{RoleOfVortices:Supp:FigS3} we show that similarly to the 280nm thick sample also for the thinner, 100nm and 26nm, films the mappings presented in Fig.~4 of the main-text also maps $I_{c}^{\perp}$ onto $I_{c}^{||}$.
Unfortunately we did not measure $I_{c}^{||}$'s of the 57nm thick film due to a technical error and the $I-V$'s of the 22nm thick film were very non-linear but did not show discontinuities.

The mapping parameters of Eq.~1 for different samples are written in Table.~\ref{RoleOfVortices:Table1}. The column of SC / INS refers to the phase where we conducted the mapping.

\begin{table} [h!]
	\vspace{1cm}
	\centering
	\begin{tabular}{|l| c | c | c | c |} 
		\hline
		Sample name & thickness [nm] & SC / INS & $B_{0}$ [T] & $a$\\ [0.5ex] 
		\hline\hline
		AD12a & 22 & SC & 4.65 & 0.71\\ 
		\hline
		AD12a & 26 & SC & 5.2 & 0.64\\ 
		\hline
		AD12a & 57 & SC & 2.4 & 0.857\\ 
		\hline
		AD12a & 100 & SC & 2 & 0.905\\ 
		\hline
		AD12a & 280 & SC & 2 & 0.91\\ 
		\hline
		RAM005b & 30 & INS & 2.72 & 1.02\\ 
		\hline
	\end{tabular}
	\caption{\bf Mapping parameters of Eq.~1 for different samples.}
	\label{RoleOfVortices:Table1}
\end{table}

\section{Mapping between $\boldsymbol{B_{\perp}}$ and $\boldsymbol{B_{||}}$ in the insulating phase of a $\boldsymbol{B}$-driven SIT}
\begin{figure*} [h!]
	\centering
	\includegraphics [height=6 cm] {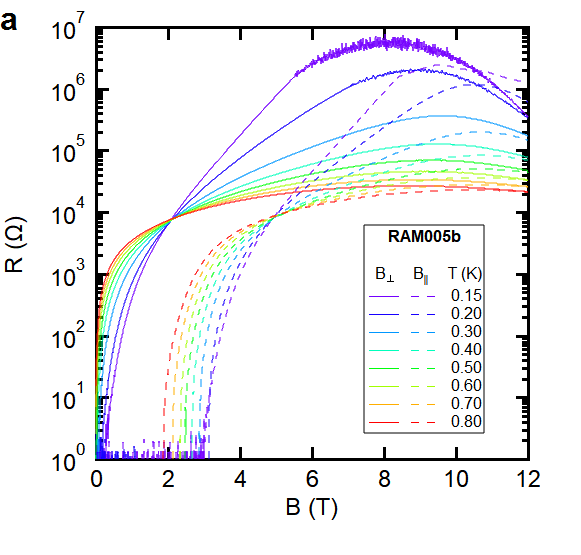}
	\includegraphics [height=6 cm] {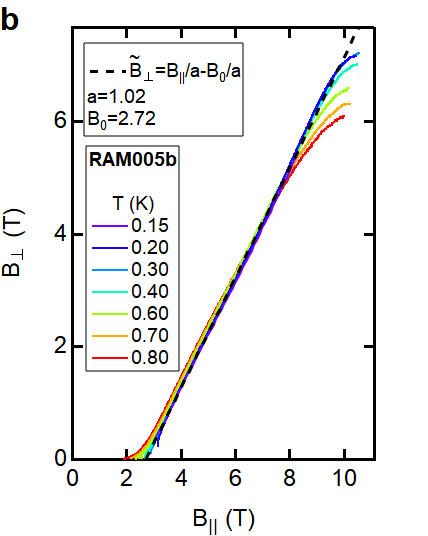}	
	\caption{{\bf Mapping of equilibrium $\boldsymbol{R(B)}$'s between $\boldsymbol{B_{\perp}}$ and $\boldsymbol{B_{||}}$ in the insulating phase of a $\boldsymbol{B}$-driven SIT.} 
		(a) $R(B_{\perp})$ (continuous lines) and $R(B_{||})$ (dashed lines) at different $T$'s.
		(b) Continuous mapping between $R$ at $B_{||}$ and $B_{\perp}$. Each point on the $B_{||}-B_{\perp}$ plane satisfies $R(B_{||},T)=R(B_{\perp},T)$. The color-coding marks different $T$'s and the dashed black line is a linear fit.
	}			
	\label{RoleOfVortices:Supp:FigS4}
\end{figure*}

In the main-text we discussed the mapping between equilibrium transport at $B_{\perp}$ and $B_{||}$ in the $B$-driven insulating phase of a highly disordered a:InO film (RAM005b). 
In Fig.~\ref{RoleOfVortices:Supp:FigS4}a we plot $R(B_{\perp})$ (continuous lines) and $R(B_{||})$ (dashed lines) at different $T$'s.
In Fig.~\ref{RoleOfVortices:Supp:FigS4}b we plot the mapping between $R$ at $B_{||}$ and $B_{\perp}$ where each point on the $B_{||}-B_{\perp}$ plane satisfies $R(B_{||},T)=R(B_{\perp},T)$
The dashed black line is a linear fit.

\begin{figure*} [h!]
	\centering
	\includegraphics [height=6 cm] {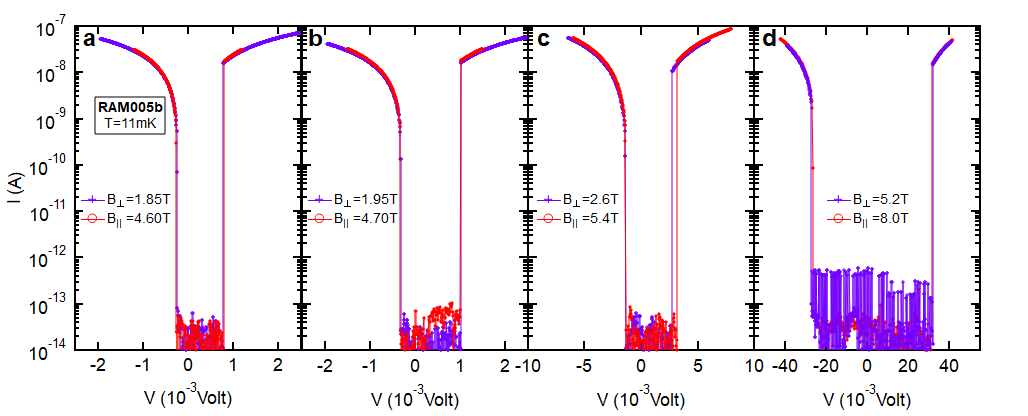}
	\caption{{\bf Mapping of Non-equilibrium $\boldsymbol{I-V}$'s.} 
		(a)-(d) $I-V$'s measured at $B_{\perp}$ (purple) and $B_{||}$ (red) at various $B$'s in the insulating phase.
	}			
	\label{RoleOfVortices:Supp:FigS5}
\end{figure*}

In Fig.~\ref{RoleOfVortices:Supp:FigS5} we plot several $I-V$'s of matching-$B$'s in different regions in the insulating phase.

\bibliographystyle{apsrev}